\documentclass[twocolumn,showpacs,amsmath,amssymb,pra]{revtex4}

\usepackage{hyperref}

\usepackage{graphicx}   	% Include figure files
\usepackage{dcolumn}    	% Align table columns on decimal point
\usepackage{bm}         	% bold math
\newcommand{\rd}{{\rm d}}
\newcommand{\ri}{{\rm i}}
\newcommand{\kL}{k_{\rm L}}
\newcommand{\Er}{E_{\rm r}}
\begin{document}

\title[Strong-field simulators]
      {Driven optical lattices as strong-field simulators}

\author{Stephan Arlinghaus}
\author{Martin Holthaus}

\affiliation{Institut f\"ur Physik, Carl von Ossietzky Universit\"at,
	D-26111 Oldenburg, Germany}

\date{June 21, 2010}

\begin{abstract}
We argue that ultracold atoms in strongly shaken optical lattices can be 
subjected to conditions similar to those experienced by electrons in 
laser-irradiated crystalline solids, but without introducing secondary
polarization effects. As a consequence one can induce nonperturbative 
multiphoton-like resonances due to the mutual penetration of ac-Stark-shifted 
Bloch bands. These phenomena can be detected with a combination of currently 
available laboratory techniques. 
\end{abstract}

\pacs{67.85.-d, 32.80.Xx, 42.50.Hz}

% 67.85.-d 	Ultracold gases, trapped gases
% 42.50.Hz 	Strong-field excitation of optical transitions in quantum 
%		systems; multiphoton processes; dynamic Stark shift 
% 32.80.Xx 	Level crossing and optical pumping

\maketitle

%%%%%%%%%%%%%%%%%%%%%%%%%%%%%%%%%%%%%%%%%%%%%%%%%%%%%%%%%%%%%%%%%%%%%%%%%%%%%%%%

\section{Introduction}
\label{S_1}

The investigation of ultracold atoms in optical lattices constitutes 
a major area of topical research~\cite{JakschZoller05,MorschOberthaler06,
LewensteinEtAl07,BlochEtAl08}. One of the long-term visions driving this
trend stems from the prospect of using these well-controllable and flexible 
systems for ``emulating'' important quantum many-body problems which still 
are not fully understood, such as high-$T_c$ 
superconductivity~\cite{HofstetterEtAl02,Bloch08}, and of obtaining information
on these by observing their cold-atom-emulated versions in the laboratory, 
rather than attempting necessarily imperfect computer simulations. So far, 
interest has been focused mainly on systems governed by a time-independent
Hamiltonian operator, a hallmark example being provided by the 
Bose-Hubbard model~\cite{GreinerEtAl02}. However, it is feasible to subject 
the lattice atoms to time-dependent external forces, and thus to study 
explicitly time-dependent phenomena~\cite{HuberEtAl07,GaulEtAl09}. 
Already in 1998 Madison {\em et al.\/} have obtained evidence for 
Bloch band narrowing with cold sodium atoms in time-periodically 
forced optical lattices~\cite{MadisonEtAl98}; more recently, dynamic 
localization~\cite{LignierEtAl07,EckardtEtAl09}, photon-assisted 
tunneling~\cite{SiasEtAl08}, and coherent control of the superfluid-to-Mott
insulator transition~\cite{ZenesiniEtAl09} have been demonstrated with 
Bose-Einstein condensates in such strongly shaken periodic potentials. 
Moreover, it has been suggested to employ oscillating optical lattices for 
realizing frustrated quantum antiferromagnetism~\cite{EckardtEtAl10}. In this 
article we argue that ultracold atoms in forced optical lattices also lend 
themselves to the study of multiphoton-like transitions under strong-field 
conditions which are barely accessible with electrons in solids irradiated 
by high-power lasers; in particular, they provide an exceptionally clean 
testing ground for the investigation of nonperturbative multiphoton-like 
resonances. We first sketch in Sec.~\ref{S_2} the required setup, and specify 
the orders of magnitude of the relevant parameters which characterize the 
optical-lattice analogs of strong laser fields. We then present numerical 
model calculations in Sec.~\ref{S_3}, demonstrating how both perturbative and 
nonperturbative resonances manifest themselves. The explanation of these
phenomena makes use of both the spatial periodicity of the optical lattice 
and the temporal periodicity of the driving force: Effectively, one encounters 
a spatiotemporal crystal, the band structure of which is controlled by the
parameters of the driving force. This viewpoint is emphasized in the
concluding Sec.~\ref{S_4}.

\section{Simulating strong laser fields}
\label{S_2}

A one-dimensional (1D) optical lattice is created, for example, by shining 
laser radiation with wavelength $\lambda = 2\pi/\kL$ against a mirror and
retroreflecting the beam into itself. An atom of mass~$M$ moving in this 
standing light wave then experiences a periodic potential with a depth~$V_0$ 
which is proportional to the laser intensity~\cite{MorschOberthaler06}. 
Mounting the mirror on a piezoelectric actuator now allows one to let it 
oscillate sinusoidally with a precisely controlled angular 
frequency~$\omega$ and amplitude~$L$, thus shaking the lattice back and 
forth~\cite{ZenesiniEtAl09}. In the laboratory frame, the Hamiltonian 
describing the particle's center-of-mass motion along the lattice direction 
then reads  
\begin{equation}
	H_{\rm lab} = \frac{p^2}{2M} 
	+ \frac{V_0}{2}\cos\left\{2\kL[x - L\cos(\omega t)]\right\} \; . 
\label{eq:LAB}
\end{equation}
The relevant characteristic energy scale is given by the single-photon 
recoil energy,
\begin{equation}
	\Er = \frac{\hbar^2\kL^2}{2M} \; ;
\end{equation}
typical scaled lattice depths $V_0/\Er$ range between about $5$ and $10$. 
For example, with $^{87}$Rb atoms in a lattice erected by light with 
wavelength $\lambda = 842$~nm one has $\Er = 1.34 \times 10^{-11}$~eV, 
as corresponding to the recoil frequency 
$\nu_{\rm r} = \Er/(2\pi\hbar) = 3.23$~kHz.

Performing a unitary transformation to a frame co-moving with the lattice, the 
Hamiltonian acquires the suggestive form~\cite{MadisonEtAl98,DreseHolthaus97} 
\begin{equation}
	H = \frac{p^2}{2M} + \frac{V_0}{2}\cos(2\kL x)
	- F x \cos(\omega t) \; ,
\end{equation}
with $F = M L \omega^2$ denoting the amplitude of the inertial force 
appearing in this oscillating frame. A meaningful measure for the strength 
of this force is the dimensionless parameter~\cite{EckardtEtAl09}
\begin{equation}
	K_0 = \frac{Fd}{\hbar\omega} \; ,
\label{eq:PAR}
\end{equation}
where $d = \lambda/2$ specifies the lattice constant. In terms of quantities
directly accessible in the laboratory, one has  
\begin{equation}
	K_0 = \frac{\pi^2}{2} \frac{\nu}{\nu_{\rm r}} \frac{L}{d} \; ,	
\end{equation}
with the driving frequency $\nu = \omega/(2\pi)$, showing that one may easily 
realize values $K_0 > 1$ when both ratios $\nu/\nu_{\rm r}$  and $L/d$ are on 
the order of unity~\cite{LignierEtAl07,EckardtEtAl09,SiasEtAl08,ZenesiniEtAl09}. 
To appreciate what this means, consider an atomic analog: A common KrF exciplex
laser provides photons with energy $\hbar\omega = 5.0$~eV. Inserting this into 
the expression~(\ref{eq:PAR}), taking the Bohr radius for the length~$d$, 
and solving for the electric field strength ${\mathcal E} = F/e$ acting on 
an electronic charge, one finds that $K_0 = 1$ is reached only for 
${\mathcal E} = 9.45 \times 10^{10}$~V/m, which is roughly one-fifth of the 
field formally experienced by a ground-state electron in the hydrogen atom. 
In this sense, time-periodically forced optical lattices can serve even as 
superstrong-field simulators: Shaking a lattice with large amplitudes~$L$ 
according to the Hamiltonian~(\ref{eq:LAB}) simulates perfectly homogeneous 
fields acting on particles in periodic potentials in the regime $K_0 > 1$ of 
the parameter~(\ref{eq:PAR}) which is hard to reach with laser-driven electrons
in traditional solids, without introducing, for example, detrimental 
polarization effects. Thus, ultracold atoms in driven optical lattices offer 
the unique possibility to study superstrong-field--induced multiphoton-like 
processes in periodic potentials in their purest form.

\begin{figure}[t]
\centering
\includegraphics[width = 0.70\linewidth]{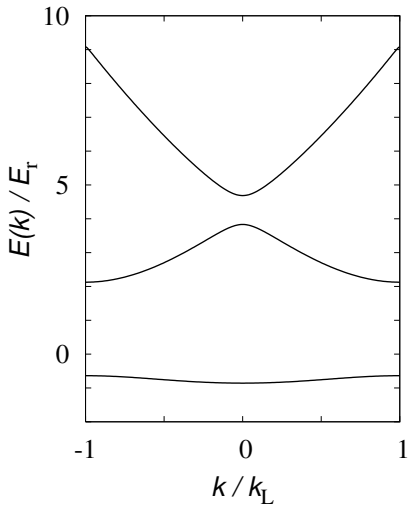}
\caption{Lowest three Bloch bands of a 1D optical lattice with depth
	$V_0 = 5.7 \, \Er$. The lowest band gap is $2.763 \, \Er$ at
	$k = \kL$.}
\label{fig:F_1}
\end{figure}

\section{Perturbative and nonperturbative multiphoton transitions}
\label{S_3}

For illustrating the dynamics that become explorable in this way, we consider 
a 1D lattice with depth $V_0 = 5.7 \, \Er$. Its single-particle eigenstates 
are Bloch waves~\cite{AshcroftMermin76},
\begin{equation}
	\varphi_{n,k}(x) = \exp(\ri k x) u_{n,k}(x) \; ,
\label{eq:BLW}
\end{equation}
with lattice-periodic functions $u_{n,k}(x) = u_{n,k}(x+d)$ labeled by 
a band index~$n$ and a wave number~$k$; Fig.~\ref{fig:F_1} depicts the 
energy dispersion relations $E_n(k)$ for the lowest bands $n = 1,2,3$. 
In the center of the Brillouin zone, that is, at $k/\kL = 0$, one has 
$E_2(0) - E_1(0) = 4.690 \, \Er$, and $E_3(0) - E_1(0) = 5.544 \, \Er$. 
We now take an initial state exclusively populating the lowest band, as 
described by
\begin{equation}
	\psi(x,t_0) = \int_{-\kL}^{+\kL} \! \rd k \, 
	g_1(k,t_0) \varphi_{1,k}(x,t_0)
\label{eq:INI}
\end{equation}        
with a Gaussian $k$-space distribution
\begin{equation}
	g_1(k,t_0) = (2\kL\sqrt{\pi}\Delta k)^{-1/2}
	\exp\left(-\frac{k^2}{2 (\Delta k)^2} \right)
\label{eq:DIS}
\end{equation}
centered around $k/\kL = 0$, and set $\Delta k = 0.1 \, \kL$ for its width, as
appropriate for an initial ensemble of noninteracting ultracold atoms. This 
state then is subjected to pulsed forcing with an amplitude $F(t)$ which rises 
from zero to a maximum value, stays constant for a while, and decreases back 
to zero. For the sake of definiteness, we consider conditions as already 
realized experimentally in Ref.~\cite{ZenesiniEtAl09}: We take $^{87}$Rb as 
atomic species in a lattice with $\lambda = 842$~nm, and design the envelope 
of the pulse such that $F(t)$ rises linearly within 10~ms, stays constant for 
a holding time of 2~ms, and then is linearly switched off in another 10~ms. 
For a driving frequency of 5~kHz, say, the ramp time of 10~ms corresponds to 
50~cycles, so that the relatively slowly changing envelope $F(t)$ may enable 
adiabatic following under nonresonant conditions.

\begin{figure}[t]
\centering
\includegraphics[width = 0.90\linewidth]{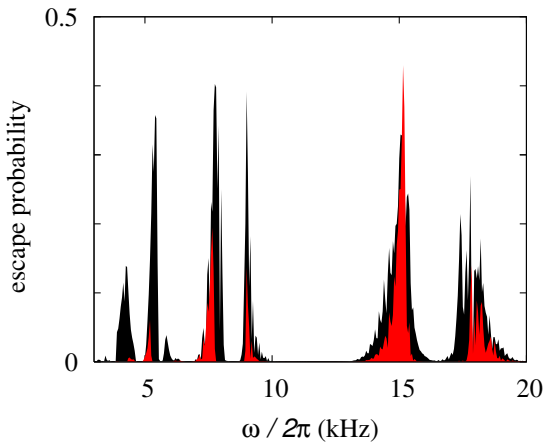}
\caption{(Color online) Escape probabilities from the lowest Bloch band after 
	pulses with linear switch-on and switch-off ramps of 10-ms duration
	each, and a holding time of 2~ms, during which a specified value 
	$K_0^{\rm max}$ of the scaled amplitude~(\ref{eq:PAR}) is reached. 
	Driving frequencies $\omega/(2\pi)$ correspond to $^{87}$Rb in an
	optical lattice with $\lambda = 842$~nm. Light, $K_0^{\rm max} = 0.7$; 
	black, $K_0^{\rm max} = 1.3$. Of particular interest is the unexpected,
	strong, and narrow resonance at $\omega/(2\pi) = 5.3$~kHz.}
\label{fig:F_2}
\end{figure}

\begin{figure}[t]
\centering
\includegraphics[width = 1.0\linewidth]{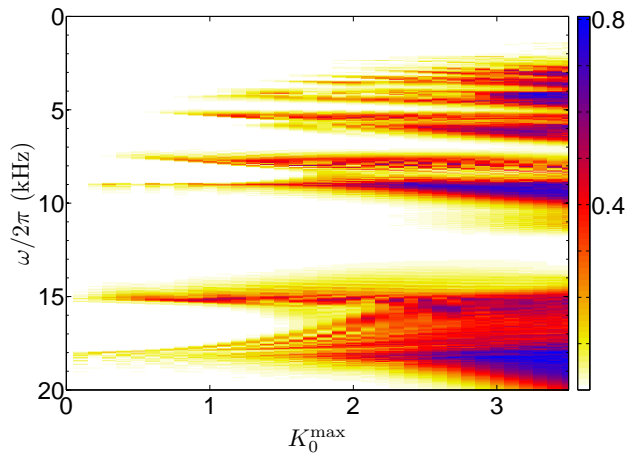}
\caption{(Color online) Escape probability versus both driving frequency 
	$\omega/(2\pi)$ and maximum scaled amplitude~$K_0^{\rm max}$, 
	for the same pulse shape as employed in Fig.~\ref{fig:F_2}.}  
\label{fig:F_3}
\end{figure}

Moreover, we rely on the fact that the fraction of atoms surviving in the 
lowest band can be accurately determined, as demonstrated by the Landau-Zener 
measurements reported in Ref.~\cite{ZenesiniEtAl09b}. We therefore compute 
the escape probability from the lowest band after each pulse, for specified 
values of $K_0^{\rm max}$ reached during the plateau phase. 
Figure~\ref{fig:F_2} shows results thus obtained for $K_0^{\rm max} = 0.7$ and
$K_0^{\rm max} = 1.3$, as functions of the driving frequency $\omega/(2\pi)$. 
The pronounced peak pattern depends markedly on the maximum driving amplitude; 
for instance, a further peak has appeared for $K_0^{\rm max} = 1.3$ at 
$\omega/(2\pi) \approx 4$~kHz which was not visible for $K_0^{\rm max} = 0.7$. 
A more complete picture is provided by Fig.~\ref{fig:F_3}, which shows a
two-dimensional plot of the escape probability considered as function of 
both $\omega/(2\pi)$ and $K_0^{\rm max}$, for the same pulse shape as taken 
in Fig.~\ref{fig:F_2}.

The positions of most of the peaks in Figs.~\ref{fig:F_2} and \ref{fig:F_3} ({\em i.e.\/}, most of the system's resonant frequencies) are easily explained: 
Because the initial state is narrowly centered around $k/\kL = 0$, its 
response is mainly determined by the energies $E_n(0)$ in the Brillouin-zone 
center. Hence, one expects ordinary $m$-photon-like resonances between the 
initial band $n = 1$ and higher bands $n = 2,3,\ldots$ when the driving 
frequency complies with the condition
\begin{equation}
	\Delta E_{n,1} \equiv E_n(0) - E_1(0) = m\hbar\omega
\label{eq:RES}
\end{equation}
for integer~$m$. Indeed, listing these expected $m$-photon transition
frequencies in Table~\ref{tab:T_1} and comparing them to the frequencies
of the peaks observed in Fig.~\ref{fig:F_3}, one generally finds quite good 
agreement.

\begin{table}
\begin{tabular}{ccccc} 
\hline
\hline
$m$ & $n$ & $\Delta E_{n,1}/(m\Er)$ & $\nu_{\rm{res}}$ (kHz) & 
$\nu_{\rm{peak}}$ (kHz)\\ 
\hline
$1$ & $3$ & $5.544$ & $17.932$ & $18.00 $ \\ %\hline
$1$ & $2$ & $4.690$ & $15.170$ & $15.15 $ \\ %\hline
$2$ & $3$ & $2.772$ & $ 8.966$ & $ 9.00 $ \\ %\hline
$2$ & $2$ & $2.345$ & $ 7.585$ & $ 7.60 $ \\ %\hline
$3$ & $3$ & $1.848$ & $ 5.977$ & $ 5.85 $ \\ %\hline
$ $ & $ $ & $  -- $ & $   -- $ & $ 5.30 $ \\ %\hline
$3$ & $2$ & $1.563$ & $ 5.057$ &    NV \  \\ 
\hline
\hline
\end{tabular}
\caption{Expected and computed resonance frequencies: $\nu_{\rm{res}}$ is 
	an $m$-photon transition frequency according to Eq.~(\ref{eq:RES}), 
	$\nu_{\rm{peak}}$ is the position of the corresponding peak where 
	it becomes apparent in Fig.~\ref{fig:F_3}. The entry NV	indicates 
	that no peak is visible for the pulse profile employed here.} 
\label{tab:T_1}
\end{table}

In some instances, however, the numerical solution of the Schr\"odinger 
equation produces a peak which does {\em not\/} fit into this naive pattern. 
Most notably, the sharp spike visible in Fig.~\ref{fig:F_2} at 
$\omega/(2\pi) = 5.3$~kHz does not match Eq.~(\ref{eq:RES}) for any 
reasonable combination of $n$ and $m$. Such ``nonperturbative'' events are 
our main concern; we predict that they can be detected experimentally in 
already existing setups. These particular resonances admit a systematic 
explanation which forces us to go way beyond the perturbative reasoning  
underlying Eq.~(\ref{eq:RES}).

\begin{figure}[t]
\centering
\includegraphics[width = 0.80\linewidth]{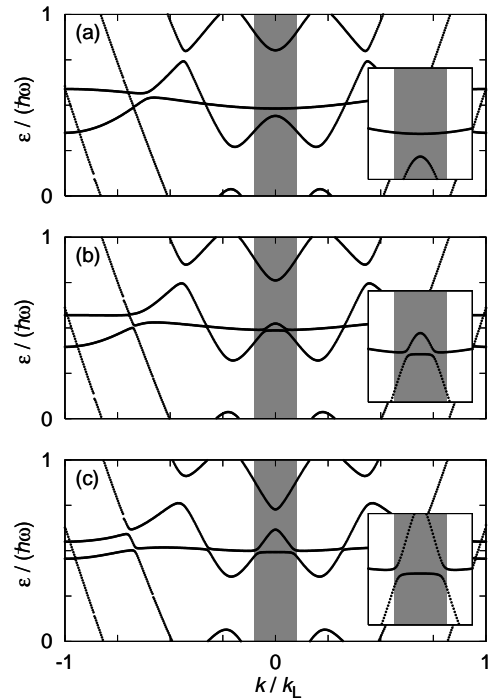}
\caption{Quasienergies $\varepsilon_n(k)$ for the 1D optical lattice driven 
	with frequency $\omega/(2\pi) = 5.30$~kHz, and scaled amplitudes 
	$K_0 = 0.7$ (a), $1.0$ (b), and $1.3$ (c). The areas shaded in gray,
	extending from $k = -0.1 \, \kL$ to $k = +0.1 \, \kL$,  
	mark the range of wave numbers explored by the initial wave packet. 
	The insets show how the quasienergy band $n = 1$ (above) is pinched 
	through with increasing~$K_0$ by the band $n=2$, displaced downward 
	by $3\hbar\omega$. This causes the nonperturbative resonance observed 
	in Fig.~\ref{fig:F_2}.}  
\label{fig:F_4}
\end{figure}

Because the Hamiltonian~(\ref{eq:LAB}) is periodic {\em both\/} in space
(with lattice period $d = \pi/\kL = \lambda/2$) {\em and\/} in time (with
driving period $T = 2\pi/\omega$), it gives rise to spatiotemporal Bloch 
waves~\cite{DreseHolthaus97},
\begin{equation}
	\psi_{n,k}(x,t) = u_{n,k}(x,t) 
	\exp\left\{\ri [k x - \varepsilon_n(k)t/\hbar]\right\},  
\label{eq:STB}
\end{equation}
with functions $u_{n,k}(x,t) = u_{n,k}(x+d,t) = u_{n,k}(x,t+T)$ reflecting
translational invariance in space and time on equal footing, 
and quasienergies $\varepsilon_n(k)$, in generalization of the usual Bloch 
waves~(\ref{eq:BLW}). While quasimomenta $\hbar k$ are determined up to an 
integer multiple of $2\pi\hbar/d = 2\hbar\kL$, quasienergies are likewise 
determined up to an integer multiple of the photon energy 
$2\pi\hbar/T = \hbar\omega$.  Figure~\ref{fig:F_4} shows one ``quasienergy 
Brillouin zone'' (of height $\hbar\omega$) with states originating from the 
lowest three Bloch bands for $\omega/(2\pi) = 5.30$~kHz, the frequency of 
the extraordinary peak in Fig.~\ref{fig:F_2}, and $K_0 = 0.7$, $1.0$, and 
$1.3$. There are various avoided crossings indicating multiphoton-like 
couplings between the bands; however, with $\Delta k = 0.1 \, \kL$ the 
wave packet evolving from the initial distribution~(\ref{eq:DIS}) mainly 
explores the interval of quasimomenta indicated by the shaded areas. The 
quasienergy band originating from the lowest unperturbed energy band $n = 1$ 
is shown enlarged in the insets; with increasing $K_0$ this band is pierced 
through from below by the quasienergy band $n=2$, displaced down in energy by 
$3\hbar\omega$ against that representative which is continuously connected to 
the bare $n=2$ Bloch band. This penetration results in ``active'' avoided 
crossings signaling a strong-field--induced three-photon resonance; this is 
responsible for the anomalous peak at $\omega/(2\pi) = 5.30$~kHz.

\begin{figure}[t]
\centering
\includegraphics[width = 0.8\linewidth]{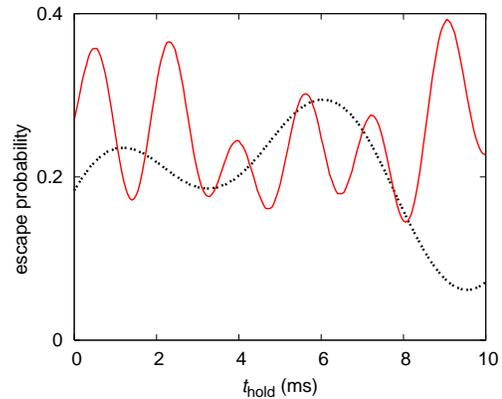}
\caption{(Color online) St\"uckelberg oscillations of the escape probability 
	in response to prolongation of the plateau duration $t_{\rm hold}$, 
	for $\omega/(2\pi) = 5.30$~kHz, and $K_0^{\rm max} = 1.0$ (dashed)
	and $1.3$ (solid line).}
\label{fig:F_5}
\end{figure}

The dynamics underlying that peak should thus be discussed in terms of the
morphology of the surfaces which emerge when the quasienergies are considered 
as functions of both the wave number~$k$ and the instantaneous amplitude~$F$ 
(or $K_0$): When the driving amplitude $F(t)$ increases during the upward 
ramp of a pulse, the initial distribution is shifted almost adiabatically 
on its quasienergy surface, parallel to the $K_0$ axis. As long as the 
maximum value of $K_0$ lies below the critical regime where this surface is 
first being pierced by another one, the initial distribution is restored 
with only minor distortion when the amplitude returns to zero, resulting 
in negligible escape probability. However, when the moving distribution 
hits an avoided-crossing regime, part of the wave function undergoes a 
Landau-Zener-type transition to the anticrossing band. Both parts of the wave 
function then evolve separately on their respective surfaces, until they meet 
for a second time during the downward ramp, when they interfere and thereby
establish the final occupation probabilities of the bands involved. This 
mechanism of splitting and interference implies that there should be 
St\"uckelberg-like oscillations when the final occupation probabilities are 
monitored while the length of the pulses' plateau segment is varied, because 
varying the plateau duration means varying the relative phase picked up by 
the two interfering components. Indeed, these oscillations are clearly visible 
in Fig.~\ref{fig:F_5}.

We remark that the standard perturbative $m$-photon resonances can be grasped 
in a similar manner: For frequencies such that Eq.~(\ref{eq:RES}) holds, two 
quasienergy surfaces are degenerate already at $F = 0$, so that adiabaticity 
is disabled and the wave function splits right at the beginning of a 
pulse~\cite{HolthausJust94}. Seen against this background, a perturbative 
resonance corresponds to the removal of a quasienergy degeneracy already 
present at $F = 0$, while a nonperturbative one emerges when ac-Stark-shifted 
Bloch bands penetrate each other at a certain finite driving strength.

\section{Conclusions}
\label{S_4}

When viewing a time-periodically forced optical lattice as a spatiotemporal
crystal, the natural basis states are the spatiotemporal Bloch
waves~(\ref{eq:STB}); the energy bands $E_n(k)$ of the undriven lattice
turn into quasienergy bands $\varepsilon_n(k)$. The latter depend not only
on the lattice parameters, but also on the parameters of the driving force. 
While they differ barely from the unperturbed energy bands as long as the 
driving amplitude is weak, corresponding to values $K_0 \ll 1 $ of the 
dimensionless para\-meter~(\ref{eq:PAR}), they become strongly distorted, 
and even penetrate each other, in the nonperturbative regime.

When subjected to pulsed forcing with an amplitude which changes slowly
compared to the period $T = 2\pi/\omega$ of the drive, a wave packet can
adjust itself adiabatically to a mere distortion of its quasienergy band.
However, when the wave packet explores a part of a quasienergy band which 
is pierced by another one, as exemplified in Fig.~\ref{fig:F_4}, Landau-Zener 
transitions occur; this mechanism leads to strong nonperturbative resonances 
at frequencies not given by the simple condition~(\ref{eq:RES}). In principle,
such resonances should also occur in solids irradiated by strong laser pulses; 
however, there they would be masked by a host of competing effects. The 
experimentally proven good controllability of ultracold atoms in forced optical
lattices makes such systems a far better testing ground for these dynamics.  
    
Our study has been restricted to the single-particle level; it is reasonable 
to expect that the phenomena exemplarily discussed in the present work can 
immediately be detected with sufficiently dilute or close-to-ideal 
Bose-Einstein condensates in driven optical lattices~\cite{EckardtEtAl09}. 
Even more, it appears equally feasible to perform the experiments suggested 
here under conditions of sizable interparticle interactions, or even of 
strong correlations. The question how the single-particle scenario outlined 
above is modified then opens up far-reaching further lines of investigation,
concerning both experiment and theory.

\begin{acknowledgments}
We thank Oliver Morsch for continuing in-depth discussions of the Pisa
experiments~\cite{LignierEtAl07,EckardtEtAl09,SiasEtAl08,ZenesiniEtAl09,
ZenesiniEtAl09b}. This work was supported by the Deutsche 
Forschungsgemeinschaft under Grant No.~Ho~1771/6. 
\end{acknowledgments}

\end{document}